\documentstyle[12pt]{article}
\advance\voffset by -1.0in
\advance\hoffset by -0.75in
\input amssym.def
\input amssym
\newcommand{\Hil}{{\cal H}}

\newcommand{\ze}{{\Bbb Z}}
\newcommand{\re}{{\Bbb R}}

\begin{document}\title{Conjectured ${Z\!\!\!\!\!\!Z}_2$-Orbifold Constructions of Self-Dual
Conformal Field Theories at Central Charge 24 - the Neighborhood Graph}
\author{P.S. Montague\\
Department of Mathematics\\
University of Adelaide\\
Adelaide\\
South Australia 5005\\
Australia}
\maketitle
\begin{abstract}
By considering constraints on the dimensions of the Lie algebra
corresponding to the weight one states of $\ze_2$ and $\ze_3$
orbifold models arising from imposing the appropriate modular
properties on the graded characters of the automorphisms on
the underlying conformal field theory, we propose a set of constructions
of all but one of the 71 self-dual meromorphic bosonic conformal
field theories at central charge 24. In the $\ze_2$ case, this
leads to an extension of the neighborhood graph of the even self-dual
lattices in 24 dimensions to conformal field theories, and we demonstrate
that the graph becomes disconnected.
\end{abstract}
\vfill
\eject
\section{Introduction}
The problem of the classification of two-dimensional conformal field theories
has seen much activity and significant progress over the last decade.
While many approaches rely on, for example, the classification of fusion rules
of some chiral algebra {\cite{Verlinde},
such methods ignore those theories for which these
fusion rules are trivial. These theories however are themselves far from trivial,
an example being the natural module for the Monster group, the so-called
Moonshine module \cite{FLMbook}.
Thus, these ``self-dual" theories must be classified separately from
this mainstream approach.

While such theories are of interest in their own right and 
serve as a simpler arena in which
to understand the general structure of conformal field theory,
they also find physical application. For example, the classification
of the self-dual theories at central charge 24 is relevant to the classification
of 10 dimensional heterotic strings \cite{Schell:Venkov}.
There is a partial classification of these theories due to Schellekens
\cite{SchellYank:curious,Schell:Venkov,SchellComplete}.
Of the 71 potential theories (under assumptions we will discuss below),
explicit constructions have only
been proposed for some 43 of them \cite{DGMtriality,SchellYank:curious,PSMorb}.
In this paper, we shall propose orbifold constructions for all but one
of the remaining theories by using results from \cite{PSMorb} and
exploiting the modular transformation properties
of the graded characters of automorphisms of orders two and three to obtain
information regarding the Lie algebraic content of any proposed orbifold model.

There is also considerable physical motivation for studying orbifold models
in general. Many of the interesting applications of conformal field theory
involve twisted fields, for example in the analysis of critical phenomena,
but particularly in the construction of physically realistic (super)string
models \cite{DFMS,stringorb1,stringorb2}.

We shall also show that our main result is a generalisation of a corresponding result
in the case of even self-dual lattices in 24 dimensions.
Such analogies between the theory of lattices and conformal
field theories, as described in \cite{PGmer,DGMtriality,thesis} and
illustrated by \cite{Schell:Venkov} (see our comments below), are
fruitful in that they enable one to make conjectures based on our understanding
of the structures at simpler levels. It is to be hoped that such investigations
will shed light on the intricate structures involved in conformal field theory.
\section{Modular constraints on dimensions}
Let $\Hil$ be a meromorphic chiral bosonic conformal field theory 
of central charge $c$ \cite{DGMtriality}
[note that from now on by the term conformal field theory we will understand
this more restricted structure]. The modes of the vertex operators of the
states of conformal
weight one in $\Hil$ form an affine Lie algebra $\widehat{g_\Hil}$ \cite{PGmer}.
Further, we say
that $\Hil$ is self-dual if its ``partition function''
\begin{equation}
\chi_\Hil(\tau)\equiv{\rm Tr}_\Hil q^{L_0-c/24}\,,
\end{equation}
where $q=e^{2\pi i\tau}$, is a modular invariant function of $\tau$.
[This is necessary for the chiral theory to be well-defined on a torus, see {\em e.g.}
\cite{Ginsparg}. For more discussion of possible distinct definitions of
self-duality see \cite{PSMtwisunique}.]
The self-dual conformal field theories can only exist at central charges
a multiple of 8. At central charges 8 and 16, the theories are easily
classified \cite{PGmer}. They are all given by the Frenkel-Kac-Segal [FKS]
construction \cite{FK,S} from an even self-dual lattice $\Lambda$
of  dimension $d$ equal to the
corresponding central charge. [Physically this describes the
propagation of a bosonic string on the torus ${\re}^d/\Lambda$.]
We denote such a conformal field theory
by $\Hil(\Lambda)$.
In 32 dimensions, the number of even self-dual lattices alone \cite{ConSlo}
is such
that an explicit enumeration is infeasible. Therefore central
charge 24 is the last possible case which may be amenable to classification.
In 24 dimensions there are 24 even self-dual lattices \cite{Venkov},
and in \cite{DGMtriality,DGMtrialsumm} it was shown that the corresponding
theories $\Hil(\Lambda)$ together with their reflection-twisted orbifolds
$\widetilde{\Hil}(\Lambda)$ (see below) give 39 inequivalent self-dual
conformal field theories of central charge 24. 
In \cite{Schell:Venkov,SchellComplete} using a beautiful
result partially
analogous to Venkov's classification of the even self-dual
lattices in 24 dimensions
\cite{Venkov}, Schellekens identified all possible
affine algebras which can correspond to a self-dual conformal field theory
at central charge 24. There are 71 such algebras
(each with distinct Lie algebra), and in each case he identified
a unique modular invariant combination of affine algebra characters.
The existence
and uniqueness of a conformal field theory corresponding to each algebra
remains to be established. Throughout this paper we shall assume uniqueness
(there are certainly no known counter-examples) and address the
problem of existence.
The main result of this paper 
is to suggest possible constructions
of most of the 71 theories and to further
reinforce the analogies with the theory
of lattices emphasized in \cite{thesis}.

We now recall the results of \cite{PSMorb}. Let $\Hil$ be a
self-dual conformal field theory at central
charge 24. The partition function for such a theory is simply of the form
$J(\tau)+c$, where $c$ is the number of states of conformal weight one
and $J$ is the classical modular function (with zero constant term).
Consider an automorphism $\theta$ of $\Hil$ of order two. We suppose that the
orbifold of $\Hil$ with respect to $\theta$ exists and is a consistent conformal
field theory (for example for the involution induced by the reflection on the
lattice, it was shown in \cite{DGMtwisted}
that the corresponding orbifold $\widetilde{\Hil}(\Lambda)$
of the (self-dual) Frenkel-Kac-Segal conformal
field theory $\Hil(\Lambda)$ constructed from an even self-dual lattice
$\Lambda$ is consistent).
In other words, if we let $\Hil_0$ denote the sub-conformal field theory \cite{DGMtriality}
invariant under $\theta$, then there exists a ``twisted representation" ${\cal K}_0$ of
$\Hil_0$ such that $\tilde\Hil=\Hil_0\oplus{\cal K}_0$ is a consistent conformal
field theory (with an obvious definition of the vertex operator structure).
We assume that this orbifold conformal field theory is also self-dual.
(Under rather natural physical assumptions, it is trivial to see that the
partition function is modular invariant - see {\em e.g.} \cite{PSMorb}.)
We write its partition function as $J(\tau)+c'$. In \cite{PSMorb}
it was shown, again under certain natural physical assumptions on the
modular transformation properties which have recently been
shown \cite{Geoff} to be equivalent to the statement
that the ground state of the twisted representation ${\cal K}$ of $\Hil$
corresponding to the representation ${\cal K}_0$ of $\Hil_0$ has conformal
weight in $\ze/2$, {\em c.f.} Vafa's ``level matching" condition \cite{Vafa}, that
\begin{equation}
\label{tempor}
c+c'=3c_0+24-24\alpha\,,
\end{equation}
where $c_0$ is the number of states of conformal weight one in $\Hil_0$
and $\alpha$ is the number of states of conformal weight $1/2$ in ${\cal K}'$.

Since the zero modes of the states of conformal weight one give the Lie algebra
$g_V$ of a conformal field theory $V$, this is simply a statement
about the relation between the dimensions of the Lie algebras corresponding
to $\Hil$, $\tilde\Hil$ and $\Hil_0$.

There is no known general procedure for writing down
the twisted sector corresponding to a given automorphism $\theta$ of an
arbitrary conformal field theory (except in the case of theories of the
form $\Hil(\Lambda)$
-- see {\em e.g.} \cite{Lepowsky,Hollthesis}, but
note the missing terms arising from normal ordering, as discussed
in \cite{DGMtwisted,PSMthird}).
However, we may use the above result to obtain a set
of possible values for the dimension of the Lie algebra of such an orbifold.
It is given by 
\begin{equation}
\label{master}
3c_0+24-c-24\alpha\,, 
\end{equation}
for some non-negative integer $\alpha$,
and must be at least $c_0$.
\section{Main method}
Now, to a given $\ze_2$-orbifold construction, there exists an inverse
$\ze_2$-orbifold construction with involution defined to be $+1$ on $\Hil_0$
and $-1$ on ${\cal K}_0$ \cite{Michael,PSMorb}. Since an automorphism of a conformal
field theory preserves the conformal weights \cite{DGMtriality}, then it
restricts to an automorphism of the corresponding Lie algebras (of order dividing
the order of the original automorphism). In other words, the Lie algebra of the
invariant theory $\Hil_0$ should be either equal to or a $\ze_2$-invariant
subalgebra of {\em both} $g_\Hil$ and $g_{\tilde\Hil}$.

The method then is to consider in turn in order of decreasing
dimension each of the algebras on Schellekens' list of
71 algebras at central charge 24 for which we already have a
constructed theory, and for each to consider all possible $\ze_2$
invariant subalgebras $g$ of the Lie algebra. We evaluate the possible
dimensions of the algebra of the orbifold theory by the above, and
then check to see whether any of the theories on Schellekens' list at
the corresponding dimension has $g$ as a $\ze_2$ invariant
subalgebra. Note that
by making the trivial observation of the existence of the orbifold
inverse, we have avoided the need to consider the $\ze_2$ invariant
subalgebras of a given Schellekens theory as sitting as an arbitrary
subalgebra inside some other Schellekens Lie algebra. Such
calculations would be complicated by the need to consider so-called
exceptional subalgebras (see, for example, \cite{Dynkin}).
It is interesting to note that in the following set of results, that
there would exist only a few spurious solutions in which the embedding
is not a $\ze_2$ invariant. We regard this as a testament to the power
of our method.

In order to reduce the size of the calculation to a more manageable
form (and avoid resort to computer calculation) we ignore potential
constructions of theories which we already have constructed, {\em
i.e.} initially just the 39 theories $\Hil(\Lambda)$ and
$\widetilde{\Hil}(\Lambda)$ for the 24 even self-dual lattices
$\Lambda$. For example, we have all of the theories
of rank 24 (in \cite{DGMtriality} it was shown that if $\rm{rank}g_\Hil=c$ then
$\Hil\cong\Hil(\Lambda)$ for some even lattice $\Lambda$, self-dual if
$\Hil$ is self-dual), and hence we need consider only those (outer) automorphisms
of $g_{\Hil(\Lambda)}$ which reduce the rank to less than or equal to 16 (the
next rank below 24 on Schellekens' list).

Once we identify a given $\ze_2$ automorphism of a given theory as
giving potentially a unique new theory, we then expand our
considerations for this automorphism to include all theories at the
appropriate dimensions, {\em i.e.} include those that we have already
constructed. If the new theory is still a unique candidate, then we
regard this as a construction, note it in Table 1, and add that theory
to our list of constructed theories.

Thus we obtain a list of conjectured constructions of theories on
Schellekens' list. Note though that to verify each construction, we
must extend the automorphism from the Lie algebra of the initial
theory to the whole conformal field theory (we make some comment on
this below in specific cases) and then verify that the orbifold theory
is consistent (along the lines of \cite{DGMtwisted}) by writing down a
set of vertex operators for a twisted sector and verifying the
appropriate locality relations hold true. This is a difficult problem,
and is left to future work.

Perhaps though the real power of this method is in showing which
constructions {\em cannot} produce consistent theories and hence restricting
effort to those which are potentially fruitful. As a simple
example, one might
consider that the common $\ze_2$ invariant subalgebra $A_4A_3U(1)$ of
${A_4}^2C_4$ and ${A_4}^2$ might indicate a possible orbifolding of
one from the other. However, the dimension of the Lie algebra of the orbifold
theory of ${A_4}^2C_4$ corresponding to an automorphism with invariant algebra
$A_4A_3U(1)$ is, by our method, $60-24\alpha$ for some non-negative
integer $\alpha$, whereas the dimension of the theory ${A_4}^2$ is 48.
Thus the orbifold theory cannot be consistent (or perhaps the automorphism
does not even lift from the Lie algebra to the conformal field theory).
In other words, the additional input obtained from considerations of modularity
properties helps to eliminate such spurious and naive solutions. One would
of course be tempted to hypothesize  that all such solutions which satisfy the
modularity constraint correspond to consistent orbifolds, though we will find
a counterexample later when we come to consider the theory ${A_2}^{12}$,
at least in so far as the automorphism may still fail to extend to the full
conformal field theory.

Once we have considered all of the 39 theories $\Hil(\Lambda)$ and
$\widetilde{\Hil}(\Lambda)$, we may then consider orbifolding the
orbifolds which we have constructed so far. In fact, this can easily
be seen to be a necessary procedure, since the rank of the orbifolded
theory (with the exception of orbifolds of the algebra $U(1)^{24}$ --
see later for a full analysis of this case) is clearly at least half
of that of the original theory. Since there is one theory on Schellekens'
list of rank 4, then we see that at least 3 successive orbifoldings from the
set of theories $\Hil(\Lambda)$ need be considered.

Having considered many of the theories in this way, towards the end we
switch techniques and look for orbifolds of the as yet unconstructed
theories in the hope of linking them by a $\ze_2$-orbifold either to
each other or to an already constructed theory.
This is obviously a more efficient technique when only a few theories
remain to be found.
\section{Results and Comments}
Obviously we will not give full details of the calculations, since
they are mainly a case by case trivial application of the above
elementary
arguments. We give a selection of examples only for illustrative
purposes, and refer the reader to Table 1 for a more detailed summary
of the results.

We first recall from \cite{Kac} the following theorem.
\vskip0.5cm
\noindent
{\bf Theorem} 
Let $g$ be a simple Lie algebra and let the (extended) Dynkin diagram
$D(g^\tau)$
of $g^{(\tau)}$, $\tau=1$, $2$, $3$, have Kac labels ${k_i}^\tau$,
$1\leq i\leq O(g,\tau)$. Let $s_0, \ldots,s_{O(g,\tau)}$ be a sequence
of non-negative relatively prime integers and set
$N=\tau\sum_{i=0}^{O(g,\tau)} {k_i}^\tau s_i$. Then the conjugacy
classes of the automorphisms of order $N$ of $g$ are in one-to-one
correspondence with the sequences $s_i$ which cannot be transformed
into one another by an automorphism of
$D(g^\tau)$. Further, the invariant Lie subalgebra is
isomorphic to the direct sum of the semi-simple Lie algebra obtained by
removing all the vertices
from $D(g^\tau)$ corresponding to non-zero $s_i$ together with a
centre $U(1)^{O(g,\tau)-r}$, where $r$ of the
$s_i$'s vanish.} 
\vskip0.5cm
Consider first the theory (with Lie algebra) $D_{24}$. [By our previous
comment regarding the rank 24 cases, this (is unique and) must be an
FKS theory.] From the ${D_{24}}^{(1)}$
and ${D_{24}}^{(2)}$ Dynkin diagrams together with the above theorem,
we see that the rank of the invariant Lie subalgebra (and hence of
any orbifold theory) is at least 23, and Schellekens' list then tells
us that the rank of the orbifold must be 24 if the theory is to be
consistent. All such theories, as discussed above, are already
constructed. Hence we have no calculation to do in this case -- all
$\ze_2$ orbifolds can only give theories of the form $\Hil(\Lambda)$
for $\Lambda$ even and self-dual. For example, the reflection twist
\cite{DGMtwisted} gives us the theory corresponding to the Niemeier
lattice ${D_{12}}^2$ \cite{DGMtriality}.

Consider now the theory $A_{17}E_7$. Again this is simply an FKS
theory. Considering the appropriate Dynkin diagrams, we see that possible
$\ze_2$ invariant subalgebras of $E_7$ are $E_6U(1)$, $A_1D_6$ and $A_7$.
We must also consider $E_7$ itself as the algebra corresponding to a $\ze_2$
invariant sub-conformal field theory in which the automorphism is
trivial on the states of weight one.
For the $A_{17}$ factor, we need only consider outer automorphisms,
since otherwise the rank of the invariant algebra will be 24 and we will
not obtain a new theory. The possible algebras that we get from outer
automorphisms are $D_9$ and $C_9$.
There are thus eight possible invariant subalgebras, all of rank 16.
At this point it is worth making the rather trivial observation that
all theories on Schellekens list have dimensions which are
a multiple of 12. In order that the dimension from $(\ref{master})$
be a multiple of 12, we require the dimension of the invariant subalgebra
to be a multiple of 4. This immediately excludes half of our eight
possible algebras, leaving
\begin{enumerate}
\item $D_9E_6U(1)$ The dimension of the ``enhanced'' ({\em i.e.} orbifold)
Lie algebra is, from $(\ref{master})$, $264-24\alpha$ and
must be at least the
dimension of the subalgebra, {\em i.e.} 232. Thus we must consider theories
with dimensions 240 or 264. At 264, there are no new theories ({\em i.e.}
theories of which we do not already have a construction). At 240, there
are new theories, but none which admit 
the invariant subalgebra as a subalgebra ($\ze_2$-invariant
or otherwise)
We conclude that this automorphism (or more
precisely the class of automorphisms with invariant algebra isomorphic
to this) does not produce any new theories by orbifolding.
\item $D_9A_7$ In this case, the upper bound on the dimension of the algebra
of the orbifold model from $(\ref{master})$ and the lower bound ({\em i.e.}
the dimension of the $\ze_2$-invariant algebra) coincide. Thus, if the
theory is to be consistent, it must have algebra isomorphic to the
invariant algebra. Such a theory exists on Schellekens' list. In fact, this
case simply corresponds to the automorphism induced by the reflection
on the lattice, {\em i.e.} the orbifold theory is $\widetilde{\Hil}(\Lambda)$,
and is known to be consistent \cite{DGMtriality}. See the comments below regarding
the special status of the reflection twisted orbifolds.
\item $C_9E_7$ As in case 1, we conclude that this class of automorphisms
cannot give rise to any new theories.
\item $C_9A_1D_6$ The ``new'' theories of dimension 288, 264 or 240
are $B_6C_{10}$, $B_5E_7F_4$ and $C_8{F_4}^2$. Only the former has
$C_9A_1D_6$ as a subalgebra ($\ze_2$-invariant or otherwise). Hence this
is a possible candidate for a new theory. To assign a greater degree of
confidence to this construction, we expand our considerations to all the theories
on Schellekens' list at the above range of dimensions. Having done this, we observe that the theory $B_6C_{10}$ remains the unique candidate.
We conjecture that this orbifold construction is consistent and produces this
theory. 
\end{enumerate}
It is worthwhile at this point
to consider the nature of the relation between this
consistency test and Vafa's ``level-matching'' condition \cite{Vafa},
{\em i.e.} that the conformal weight of the twisted sector ground
state is a half-integer. [In fact not only should we check in which
cases the possible consistency of an orbifold theory is agreed upon,
but we may also check that cases in which the conformal weight of the ground
state in the twisted sector is at least 1 correspond to the cases $\alpha=0$,
{\em i.e.} no states at conformal weight $1/2$ in the twisted sector.]
As mentioned earlier, it has been shown \cite{Geoff} that the
level-matching condition in the $\ze_2$ case is equivalent to the
assumption that the character of the automorphism transforms
appropriately under modular transformations ({\em i.e.} that it is a
$\Gamma_0(2)$ invariant).  However, our test is not {\em equivalent} to
level-matching, since we also use non-trivial input by checking potential
theories against Schellekens' list. For example, considering the invariant
algebra ${E_8}^2E_7A_1$ of ${E_8}^3$, the conformal weight of the twisted sector
ground state should, by standard results \cite{Mythesis}, be $1/4$ and so the
orbifold should be inconsistent. However, our test here cannot exclude
the possibility that the orbifold theory is ${E_8}^3$ again. On the other hand,
take the invariant algebra ${A_5}^4{A_1}^4$ of ${E_6}^4$. According to the
considerations here, we find that it cannot give a consistent orbifold
theory. However, from \cite{Mythesis} we see that the twisted sector
ground state should have conformal weight $1$. 

In general though, we conjecture that a $\ze_2$ orbifold of a self-dual theory
is consistent if and only if the ground state of the twisted sector
has conformal weight a half-integer. The apparent counterexample in the
case of ${E_6}^4$ immediately above we attribute to
the fact that the automorphism under consideration does not lift to the
conformal field theory as an involution (it is easy to see that any
inner automorphism of the Lie algebra of a FKS theory lifts to the whole
theory (for an outer automorphism one may encounter problems with the glue
vectors \cite{ConSlo}), but the appropriate definition on the cocycles
\cite{DGMtwisted} may lead to a change in the order of the automorphism
in its action on the full conformal field theory). Note that a restriction
such as the phrase
``self-dual'' is necessary, since in the case of a reflection-twist
$\widetilde{\Hil}(\Lambda)$ of an FKS theory $\Hil(\Lambda)$ the orbifold
is consistent if and only if $\sqrt 2\Lambda^*$ is an even lattice
\cite{thesis}, but the
conformal weight of the twisted sector ground state is a half-integer
for all even lattices $\Lambda$.

Our proposed test therefore, while perhaps not strictly stronger
than simple level-matching, can obviate the need to consider the
extension  or otherwise of automorphisms to the full conformal field theory,
and in cases as we will consider below in which the calculation of the conformal
weight of the twisted sector ground state is impossible, it is the only
indicator for consistency of the orbifold which we have available.
We rely on it in such cases, since in those other cases for which we can
check level-matching ({\em e.g.} the FKS theories) we find very few
orbifold theories which fail to be consistent on the grounds of 
``level-matching'' and yet appear consistent to our checks. Almost
all examples, as the ${E_8}^3$ case above, involve potential orbifolding
back to the original theory
(in which case our test is weakened since
the invariant subalgebra is then automatically a $\ze_2$-invariant
subalgebra of the orbifold algebra),
and so are irrelevant in any case in our
attempted construction of the entire list of self-dual theories at
central charge 24.

Having said that, we consider below
an example which patently fails ``level-matching''
and yet cannot be excluded by our test. It is an example of the potential
failure due to spurious solutions at low dimension which are possible.
We can exclude it either by level-matching, or, as we discuss below, by
showing that the automorphism does not extend to the conformal field theory.
In general, one must therefore be aware that a range of techniques
need be employed in order to eliminate apparently possible orbifold theories.
Ultimately, all such proposed constructions must be demonstrated
by explicit formulation of the appropriate vertex operators and 
calculation of locality
relations. The current work, as already mentioned, is in many ways simply
filtering out from the mass of possible constructions a handful of cases
which will be likely to be consistent. At the very least, we are
able to rigorously discard without excessive calculation
cases which cannot be consistent.

Consider the involution of the Lie algebra ${A_2}^{12}$ given by six
interchanges, with invariant subalgebra ${A_2}^6$. This cannot give
rise to a consistent orbifold by level-matching, since the conformal
weight of the twisted sector ground state should be $3/4$. But
according to our method, the theory ${A_2}^6$ with $\alpha=1$ is a
possible (in fact the only) candidate. There is immediate evidence
however that this is an accidental solution. The ``enhanced'' algebra
(that of the orbifold theory) is equal to the invariant algebra, and
so there are no states of conformal weight one in the putative twisted
sector. However, $\alpha=1$ indicates that there is a single state at
conformal weight $1/2$. These facts are difficult to reconcile.  It
might be thought that explicit calculation of $\alpha$ (possible in
these FKS theories) could eliminate this theory. But in fact we find
we cannot even extend the automorphism from the Lie algebra up to the
full conformal field theory. The glue code for the Niemeier lattice
${A_2}^{12}$ is the ternary Golay code. We are required to find six disjoint
transpositions in this code which leave it invariant (or equivalently,
using self-duality of the code, map each basis codeword to a codeword in the dual).
No such set of transpositions exists, and so the automorphism fails
to extend to the conformal field theory.

Since the construction of ${A_2}^4D_4$ from ${A_2}^{12}$ (see Table 1)
is also potentially a spurious low dimension solution, we can explicitly check
in that case that the automorphism does extend to the whole theory.
The automorphism which reduces $A_2$ to $A_1$ maps the glue code $1$
to $2$ (in the notation of \cite{ConSlo}). We find an appropriate symmetry of the ternary Golay code.
Note though that the definition of the action of the automorphism
on the cocycles could potentially lead to a doubling in the order
of the automorphism. Nevertheless, we feel that the checks performed
are sufficient to accept this orbifold construction as valid.

We then proceed to consider $\ze_2$ orbifolds of the theories constructed
so far ({\em i.e.} those from the FKS theories in Table 1 together
with the 15 of the
theories $\widetilde{\Hil}(\Lambda)$ distinct from the FKS theories). As mentioned above,
calculation of the twisted sector ({\em i.e.} of $\alpha$ and verification
of level-matching) is not possible in these cases, since too little  
is known about the structure of the original theories. But we rely on
our technique given its proven reliability in cases when cross-checks may be made.
The results are as indicated in Table 1.

\begin{table}
\begin{tabular}{|c|c|c||c|c|c|}
\hline
orbifold & original & invariant & orbifold & original & invariant\\
algebra & algebra & algebra & algebra & algebra & algebra\\
\hline
& & & & & \\
$E_8B_8$ & ${E_8}^3$ & $E_8D_8$ & ${A_2}^2{A_5}^2C_2$ & ${A_5}^4D_4$ & $A_1{A_2}^2
A_3A_5C_2U(1)$\\
& & & & & \\
$B_6C_{10}$ & $A_{17}E_7$ & $A_1C_9D_6$ & ${A_4}^2C_4$ & ${A_4}^6$ & ${A_4}^2{C_2}^2$ \\
& & & & & \\
$B_5E_7F_4$ & $D_{10}{E_7}^2$ & $B_4B_5E_7$ & ${A_2}^4D_4$ & ${A_2}^{12}$ & ${A_2}^4
{A_1}^4$\\
& & & & & \\
$C_8{F_4}^2$ & $A_{15}D_9$ & ${B_4}^2C_8$ & $A_3C_7$ & $B_4{C_6}^2$ & $A_1A_3C_6$ \\
& & & & & \\
$B_4{C_6}^2$ & $A_{11}D_7E_6$ & $B_2B_4C_4C_6$ & $A_1{A_3}^3$ & ${A_2}^2{A_5}^2C_2$ &
${A_1}^3{A_3}^2U(1)$\\
& & & & & \\
$A_5C_5E_6$ & ${E_6}^4$ & $A_1A_5C_4E_6$ & $A_2C_2E_6$ & ${A_3}^2{D_5}^2$ & $A_2C_2D_5U(1)$\\
& & & & & \\
$A_4A_9B_3$ & ${A_8}^3$ & $A_3A_4A_8U(1)$ & ${A_1}^2C_3D_5$ & ${C_2}^4{D_4}^2$ &
${A_1}^3C_2D_4U(1)$ \\
& & & & & \\
$A_8F_4$ & ${A_8}^3$ & $A_8B_4$ & ${A_1}^3A_7$ & ${A_2}^2{A_5}^2C_2$ & ${A_1}^4A_5U(1)$ \\
& & & & & \\
$A_3A_7{C_3}^2$ & ${A_7}^2{D_5}^2$ & ${A_1}^2A_3A_7{B_2}^2$ & ${A_2}^2A_8$ & ${A_4}^2C_4$
& $A_1A_2A_3A_4U(1)^2$\\
& & & & & \\
\hline
\end{tabular}
\caption{Conjectured orbifold constructions of 18 new theories}
\end{table}

We conclude this section with some trivial observations.

Firstly, we note that, in order that the difference between our upper
and lower bounds on the dimension of the orbifold Lie algebra be non-negative,
we require that the dimension of the invariant Lie algebra, $c_0$, be
at least ${1\over 2}(c-24)$. This value is attained for the reflection twist
on the lattice. We may ask whether the converse is true, {\em i.e} whether
this is the only involution for which
our bounds coincide.
Checking all possible Lie algebra involutions,
we find that for any simple Lie algebra the dimension of the invariant
subalgebra is always at least ${1\over 2}(d-r)$, where $d$ is the dimension
of the Lie algebra and $r$ is its rank, and further we have equality for
a unique involution in each case. Since the Lie algebra for each of the
FKS theories $\Hil(\Lambda)$ for $\Lambda$ Niemeier is of rank 24,
then we see that we have equality of our upper and lower bounds
if and only if the involution is simply the lift of the reflection twist. 
Thus, the reflection-twisted orbifolds $\widetilde{\Hil}(\Lambda)$
of the Niemeier lattices $\Lambda$ are in some sense the extreme cases
which saturate the bounds obtained from modularity constraints.

Secondly, we note that all
of the constructions which we have proposed (and which we will propose
in the section below) correspond to $\alpha=0$, {\em i.e.} to the ground
state in the twisted sector having conformal weight at least 1. We know of
no reason why this should be so, since it is not necessarily always true
as the simple example of the orbifolding of the ${E_8}^3$ theory
with an involution which is simply the reflection twist on one $E_8$
factor and the identity on the other two. This gives us back the theory
${E_8}^3$, and corresponds to $\alpha=16$. We do not believe that it is an
artifact of our method of search [unlike the fact that
most of the constructions seem to relate a given theory to
an orbifold theory with algebra of strictly smaller rank --
this is merely a result of us beginning with the rank 24 theories
and working downwards, and indeed because of the existence of an inverse
to any given orbifold construction rank reduction and rank enhancement
should obviously be equally common] and so remains to be explained.

In fact, for $\alpha=0$ or 1, the graded trace of the involution is a
hauptmodul. It is an interesting aside, in relation to the
Moonshine conjectures of Conway and Norton \cite{moonshine},
to note that  our method shows (in all but one case -- which
cannot be decided unambiguously) that
the ``triality involution'' \cite{DGMtriality,FLMbook}
(one of the few non-trivial cases of an involution known rigorously
to be well-defined on a non-FKS conformal field theory)
corresponds to a hauptmodul, assuming the relevant orbifold to be
consistent.
\section{The Neighborhood Graph}
At this stage, 14 of the 71 Schellekens theories still remain to be constructed.
We
now, as discussed earlier, and using the fact that to each orbifold
there corresponds an inverse orbifold construction, consider
orbifolding in turn each of these theories and compare, using our
techniques, against the {\em full} list of Schellekens' theories
constructed so far.
We draw a
graph (Figure 1) with solid arrows indicating the construction is the
unique possibility and dashed arrows indicating that there is an
ambiguity.
Note that an ambiguity may often be resolved by considering
orbifolding the theory at the opposing end of the corresponding dashed
arrow. We have
removed all such ambiguities (and are implicitly assuming that the
class of automorphisms with a given invariant algebra produce isomorphic
orbifold theories). Theories which have already been constructed in the above
are enclosed in a box. We have not included all arrows coming out of a theory
when there are sufficient arrows present to connect that theory (indirectly)
to a theory already constructed. The appropriate invariant algebras
are indicated on the connecting arrows.

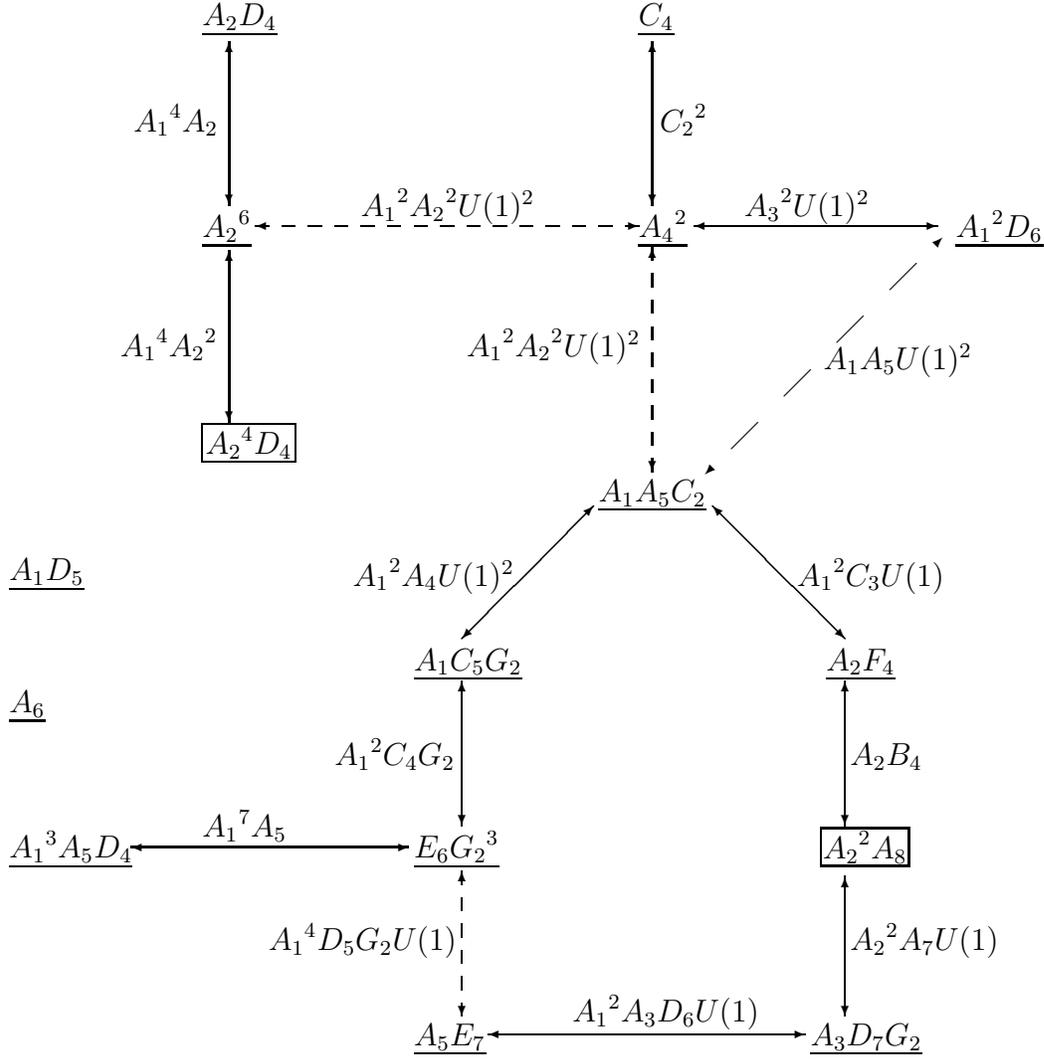
\begin{figure}
\begin{picture}(200,400)(-250,-230)
\put(0,0){\underline{$A_1A_5C_2$}}
\put(-2,-2){\vector(-1,-1){50}}
\put(-52,-52){\vector(1,1){50}}
\put(-70,-64){\underline{$A_1C_5G_2$}}
\put(43,-2){\vector(1,-1){50}}
\put(93,-52){\vector(-1,1){50}}
\put(86,-64){\underline{$A_2F_4$}}
\put(75,-32){${A_1}^2C_3U(1)$}
\put(-93,-32){${A_1}^2A_4U(1)^2$}
\put(-70,-135){\underline{$E_6{G_2}^3$}}
\put(-52,-68){\vector(0,-1){55}}
\put(-52,-123){\vector(0,1){55}}
\put(93,-68){\vector(0,-1){55}}
\put(93,-123){\vector(0,1){55}}
\put(84,-138){\framebox(33,14){${A_2}^2A_8$}}
\put(-100,-100){${A_1}^2C_4G_2$}
\put(95,-100){$A_2B_4$}
\put(-70,-206){\underline{$A_5E_7$}}
\put(80,-206){\underline{$A_3D_7G_2$}}
\put(-52,-140){\line(0,-1){5}}
\put(-52,-150){\line(0,-1){5}}
\put(-52,-160){\line(0,-1){5}}
\put(-52,-170){\line(0,-1){5}}
\put(-52,-180){\line(0,-1){5}}
\put(-52,-190){\vector(0,-1){5}}
\put(-52,-195){\line(0,1){5}}
\put(-52,-185){\line(0,1){5}}
\put(-52,-175){\line(0,1){5}}
\put(-52,-165){\line(0,1){5}}
\put(-52,-155){\line(0,1){5}}
\put(-52,-145){\vector(0,1){5}}
\put(93,-142){\vector(0,-1){53}}
\put(93,-195){\vector(0,1){53}}
\put(-42,-202){\vector(1,0){120}}
\put(78,-202){\vector(-1,0){120}}
\put(95,-170){${A_2}^2A_7U(1)$}
\put(-125,-170){${A_1}^4D_5G_2U(1)$}
\put(-10,-197){${A_1}^2A_3D_6U(1)$}
\put(-72,-131){\vector(-1,0){105}}
\put(-177,-131){\vector(1,0){105}}
\put(-223,-135){\underline{${A_1}^3A_5D_4$}}
\put(-150,-127){${A_1}^7A_5$}
\put(15,100){\underline{${A_4}^2$}}
\put(-90,108){${A_1}^2{A_2}^2U(1)^2$}
\put(55,108){${A_3}^2U(1)^2$}
\put(15,180){\underline{$C_4$}}
\put(85,50){$A_1A_5U(1)^2$}
\put(-150,15){\framebox(35,14){${A_2}^4D_4$}}
\put(-150,100){\underline{${A_2}^6$}}
\put(-150,180){\underline{$A_2D_4$}}
\put(135,100){\underline{${A_1}^2D_6$}}
\put(20,16){\vector(0,-1){5}}
\put(20,21){\line(0,1){5}}
\put(20,31){\line(0,1){5}}
\put(20,41){\line(0,1){5}}
\put(20,51){\line(0,1){5}}
\put(20,61){\line(0,1){5}}
\put(20,71){\line(0,1){5}}
\put(20,81){\line(0,1){5}}
\put(20,91){\vector(0,1){5}}
\put(20,112){\vector(0,1){62}}
\put(23,140){${C_2}^2$}
\put(-50,55){${A_1}^2{A_2}^2U(1)^2$}
\put(20,174){\vector(0,-1){62}}
\put(-140,30){\vector(0,1){65}}
\put(-140,95){\vector(0,-1){65}}
\put(-140,112){\vector(0,1){62}}
\put(-140,174){\vector(0,-1){62}}
\put(-176,140){${A_1}^4A_2$}
\put(-181,55){${A_1}^4{A_2}^2$}
\put(-125,104){\vector(-1,0){5}}
\put(-120,104){\line(1,0){5}}
\put(-110,104){\line(1,0){5}}
\put(-100,104){\line(1,0){5}}
\put(-90,104){\line(1,0){5}}
\put(-80,104){\line(1,0){5}}
\put(-70,104){\line(1,0){5}}
\put(-60,104){\line(1,0){5}}
\put(-50,104){\line(1,0){5}}
\put(-40,104){\line(1,0){5}}
\put(-30,104){\line(1,0){5}}
\put(-20,104){\line(1,0){5}}
\put(-10,104){\line(1,0){5}}
\put(0,104){\line(1,0){5}}
\put(10,104){\vector(1,0){5}}
\put(36,104){\vector(1,0){92}}
\put(128,104){\vector(-1,0){92}}
\put(45,15){\vector(-1,-1){5}}
\put(50,20){\line(1,1){10}}
\put(70,40){\line(1,1){10}}
\put(90,60){\line(1,1){10}}
\put(110,80){\line(1,1){10}}
\put(125,95){\vector(1,1){5}}
\put(-223,-80){\underline{$A_6$}}
\put(-223,-30){\underline{$A_1D_5$}}
\end{picture}
\caption{Subgraph of the ``neighborhood graph'' of the central charge 24
self-dual conformal field theories}
\end{figure}

We thus see that two of the theories, namely those with algebras
$A_1D_5$ and $A_6$, cannot be obtained by $\ze_2$ orbifolding, while
for another three theories, $C_4$, ${A_4}^2$ and ${A_1}^2D_6$,
there is some ambiguity as to whether they may be obtained.
Note that this isolation of the theories $A_1D_5$ and $A_6$ is rigorous
(whereas our conjectured constructions of course remain to be verified
in detail).

We now relate this picture to the ``neighborhood graph'' for even
self-dual lattices as described by Borcherds in \cite{ConSlo}.  We
shall show that the complete graph of all 71 self-dual central charge
24 conformal field theories with connections corresponding to
$\ze_2$-orbifold constructions of one theory from another has the
neighborhood graph of the 24 dimensional even self-dual lattices as a
sub-graph (identifying the node for the lattice $\Lambda$ with that
for the conformal field theory $\Hil(\Lambda)$, both of which in any
case we label by the (identical) corresponding Lie algebra).  The main
results of this paper can be interpreted as simply extending this
neighborhood graph. In 8 and 16 dimensions, the neighborhood graphs of
both the lattices and conformal field theories coincide.  However,
this is obviously no longer the case in 24 dimensions, and in
particular, in the light of the above comments regarding the theories
with algebras $A_6$ and $A_1D_5$, we see that the neighborhood graph
of the central charge 24 self-dual conformal field theories is
disconnected, in contrast to that of the 24 dimensional even self-dual
lattices. The framework of the conformal field theories
admits a richer structure than that of the lattices.

We begin by recalling the definition of neighboring lattices
\cite{ConSlo}.  Two lattices $A$ and $B$ are said to be neighbors if
their intersection has index two in each of them. We restrict attention to
unimodular lattices. Choose $x\in {1\over 2}A-A$ such that $x^2$ is
integral. Define $A_x=\{a\in A|a\cdot x\in\ze\}$. Then the lattice
$B=<A_x,x>$ is a unimodular neighbor of $A$. Further, all neighbors
of $A$ arise in this way. The neighborhood graph of the even self-dual
lattices in 24 dimensions is then simply given by joining vertices
corresponding to lattices by an edge if they are neighbors (note that
in this case $x^2$ is even). 

The notion of neighboring lattices is clearly analogous to the existence
of a $\ze_2$-orbifold construction between conformal field theories.
(See \cite{PGmer,thesis} for discussion of the analogies and deeper
connections which exist between the theory of lattices and conformal
field theory.) Let us make this analogy clearer. Consider two even
self-dual neighboring lattices $A$ and $B$ ($B$ constructed from $A$
as above). Consider the corresponding FKS conformal field theories
$\Hil(A)$ and $\Hil(B)$. Define an involution on $\Hil(A)$
by $\theta_A=e^{2\pi i x\cdot p}$, where $p$ is the momentum operator
on $\Hil(A)$.
Define also an involution on $\Hil(B)$ by $\theta_B|\lambda\rangle=
|\lambda\rangle$ for $\lambda\in A_x$ and  $\theta_B|\lambda\rangle=
-|\lambda\rangle$ for $\lambda\in x+A_x$ (and with trivial action
on the bosonic creation and annihilation operators).
Then the corresponding invariant sub-conformal field theories are
both trivially seen to be isomorphic to $\Hil(A_x)$. In other
words, the invariant Lie algebra under the automorphism $\theta_B$
on $\Hil(B)$ or $\theta_A$ on $\Hil(A)$ is that with root lattice
given by the span of the length squared two vectors in $A_x$ \cite{PGmer}.
It is clear in this case that the corresponding orbifold theory is
consistent, {\em i.e.} to convince oneself that the orbifold
of $\Hil(A)$ with respect to $\theta_A$ is $\Hil(B)$. Hence we must
have the relation $(\ref{tempor}$) between the dimensions of the
corresponding Lie algebras.
In this case, the dimension of the Lie algebra of $\Hil(A)$ is simply
$|A(2)|+24=24(h(A)+1)$, where $h(A)$ is the corresponding Coxeter number.
Similarly for the theory $\Hil(B)$. The dimension of the invariant
Lie subalgebra is $|A_x(2)|+24$. Consider the odd unimodular lattice
$C=A_x\cup((A-A_x)+x)$ corresponding to the neighborhood graph
link \cite{ConSlo}. Any element
of $(A-A_x)+x$ has odd square (since $x^2$ is even), and so
$|C(2)|=|A_x(2)|$, and we deduce from ($\ref{tempor}$) that
\begin{equation}
|C(2)|=8(h(A)+h(B))-16-8\alpha\,,
\end{equation}
where $\alpha$ is the number of weight one half states
in the twisted sector, and so in this case is simply the number
of vectors of length squared one in $C$. Since these lattices $C$ have minimal
norm 2 \cite{ConSlo} then $\alpha$ vanishes, and we recover the relation
given in \cite{ConSlo}.
Conversely, we can regard ($\ref{tempor}$) as being the generalization
of this result from the case of neighboring even lattices to
$\ze_2$-orbifolds of conformal field theories, thus
strengthening the analogies between the theory of lattices and conformal
field theory pursued in \cite{PGmer} and \cite{thesis} and also
exemplified in \cite{Schell:Venkov}.

We make some simple remarks. Firstly we note that it is not all
of the neighbors of an even lattice are themselves even (corresponding
to the vector $x^2$ being odd rather than even).
The analog for conformal field theories would be that the
orbifold conformal field theory is not consistent (as a bosonic theory),
in particular
in the sense that some of the relevant locality relations become
fermionic in character.

Secondly we note that the analog of the
enumeration of the 24 dimensional odd
unimodular lattices in \cite{ConSlo} by the links on the neighborhood graph
(including self-links) in the context of conformal field theory would
be the association of the self-dual super vertex operator algebra
$\Hil_0\oplus{\cal K}_1$ (where ${\cal K}={\cal K}_0\oplus{\cal K}_1$ in
the notation introduced earlier, and we decompose $\Hil$ similarly)
with a given $\ze_2$ orbifold construction. We see the conformal field theory,
the orbifold and the super vertex operator algebra $\Hil_s$
in the following picture
as the row, column and diagonal through $\Hil_0$ respectively.
\begin{equation}
\begin{array}{cccccc}
& \tilde\Hil & & \\
& \| & & \\
\Hil= & \Hil_0 & \oplus & \Hil_1 & \\
& \oplus & \oplus & & \\
{\cal K}= & {\cal K}_0 & \oplus & {\cal K}_1 & \\
& & & & \begin{picture}(10,10)
\put(10,0){\line(-1,1){10}}
\put(8,-2){\line(-1,1){10}}
\end{picture}
 \\
& & & & & \Hil_s
\end{array}
\end{equation}
The theory $\Hil\oplus{\cal K}$ is to be interpreted as the ``abelian intertwining
algebra'' containing all the above theories \cite{HuangNonMer}, and is the
``dual'' of $\Hil_0$ \cite{PGmer} in some natural sense.

Finally we note that the absence of any discernible regularity in the
pattern of the neighborhood graph in the lattice case makes 
the existence of
any such pattern
in the conformal field theory case unlikely, and indeed we see no such pattern
emerging from our (admittedly incomplete) results.

We may also remark that from the fact that
the neighborhood graph for the Niemeier lattices
is connected, that it is a subgraph of the neighborhood graph for the
conformal field theories and from our results in the preceding section,
we see that we may construct at most 69 and at least 66 (under the assumption
that when there is a unique candidate in our test then the construction
is consistent) of the 71 theories starting from any given one. In other words,
perhaps all but two of the central charge 24 self-dual conformal field
theories may be obtained by repeated $\ze_2$-orbifolding of the Moonshine
module using the constructions identified in Figure 1 and Table 1.
\section{Extension to Third Order Twists}
Let us briefly consider the extension of our method to third order twists
in order to identify possible constructions of the theories not constructed
in the $\ze_2$ analysis above.

Suitably rewriting the results of \cite{PSMorb}, we find that the analog
of ($\ref{tempor}$) is
\begin{equation}
c+c'=4c_0+24-108\alpha_1-36\alpha_2\,,
\end{equation}
where $\alpha_i$ is the number of states of conformal weight $i/3$ in the
twisted sector (and so we assume them to be non-negative integers).

We omit the details, but we find that we obtain (as the unique candidate
for a given $\ze_3$-invariant subalgebra) ${A_1}^3A_7$ from ${A_4}^2$
and $A_2F_4$ from $A_6$. However, the theory $A_1D_5$ remains isolated.

Similar relations hold true for twists of order 5 and 7, but we do not
carry through the analysis as we feel that with such high order twists
on low dimensional algebras, the possibility of accidental spurious
solutions is too great to ignore.
\section{Conclusions}
We have obtained an upper bound on the dimension of the Lie algebra of
an orbifold theory corresponding to an automorphism of order two or three
from modularity considerations. This serves
to limit the degree of enhancement of the common $\ze_2$-invariant
algebra by the twisted sectors in the orbifold model, and so aids
in many cases in the identification of a unique orbifold
candidate on Schellekens' list of 71
self-dual conformal field theories at central charge 24,
or alternatively the absence of a suitable candidate
indicates that the given orbifold cannot be consistent. 

We have thus conjectured constructions of all but one of the 71
theories.
Though the definition of the automorphism on a given conformal field theory
is not fully specified by its action on the corresponding Lie algebra,
and in fact in many cases the automorphism does not lift to
the full theory (or at least does not lift
to an automorphism of the same order), at the very
least we have narrowed down the number of cases which need be considered
by more
explicit and tedious methods
to a handful of possible constructions. The fact that the majority of the
theories appear to be given simply by successive $\ze_2$-orbifolding of a
particular theory,
such as the Moonshine module, should prove useful in many applications
and calculations.

In addition, we have
demonstrated the extension of the ``neighborhood graph" of the
even self-dual lattices to the self-dual conformal field theories
and further
demonstrated that our modular constraint may be regarded as
the analog of the relation between the dimensions of the root
lattices of neighboring even lattices and the corresponding odd
unimodular lattice, thus extending the often surprisingly deep
connections and analogies between the theory of lattices
and conformal field theory suggested in \cite{PGmer}
and continued in \cite{DGMtriality,thesis,Schell:Venkov}.

In any case, our method can be used
to rather simply, using elementary Lie algebra techniques,
exclude proposed orbifold constructions. In particular,
it can be used as a supplement to Vafa's ``level-matching" condition,
and as a replacement for the latter in cases where too little is known
of the explicit structure of the conformal field theory or how to
construct its orbifold. 
In this vein, we have demonstrated rigorously the isolation under
$\ze_2$ orbifoldings of at least 2 of the 71 self-dual $c=24$ conformal
field theories, and the isolation under both $\ze_2$ and $\ze_3$
orbifoldings of the theory with algebra $A_1D_5$.

The explicit construction of the self-dual theories by methods analogous
to those of \cite{DGMtwisted} or by the more indirect but more
powerful approach of \cite{PSMIntertwiners} will be pursued in future work.

\end{document}